# Analysis of Quantum Light Memory in Atomic Systems

H. Kaatuzian [a], A. Rostami [b], and A. Ajdarzadeh Oskouei *[a]

a) E. Eng. Dept., AmirKabir University of Technology, Tehran, Iran
b) Faculty of Electrical Engineering, Tabriz University, Tabriz 51664, Iran.
Tel: +98 411 3393724
Fax: +98 411 3300819
E-Mail: rostami@tabrizu.ac.ir

**Abstract-** We extend the theory to describe the quantum light memory in $\Lambda$ type atoms with considering $\gamma_{bc}$ (lower levels coherency decay rate) and detuning for the probe and the control fields. We obtain that with considering these parameters, group velocity of the probe pulse does not tend to zero by turning off the control field. We show that there are considerable decay for the probe pulse and the stored information. Also, it is inferred that the light field does not tend to zero when the control field is turned off. In addition, we obtain that in the off-resonance case there is considerable distortion of the output light pulse (with fast oscillations) that destroys the stored information. Then we present the limitations on detuning (bandwidths) of the probe and the control fields to achieve negligible distortion. We finally present the numerical calculations and compare them with obtained analytical results.

## I. Introduction

Atomic coherence and related phenomena such as electromagnetically induced transparency (EIT) and slow light have been studied extensively in recent years [1-10]. Many application are proposed to this phenomena such as nonlinear optics (SBS, FWM and etc.), Lasing without Inversion, Laser cooling and Sagnac Interferometer [11-14]. One of important and promising applications in this field is light storage and quantum light memory that is investigated by some research groups [15-25]. The most common mechanism in this application is that the light pulse is trapped and stored in atomic excitations in the EIT medium by turning off the control field and then is released by turning on the control field. However, most of these works do not present clearly and general theory to analyze the propagation and storage of light. In addition, most of the works in EIT and slow light treat the light classically that is not proper to extend to quantum memory in which quantum state of light is to be stored. The most general theory for quantum memory was developed by M. Fleischhauer, et. al. [16,17]. They consider the light, quantum mechanically and present an excellent theory to describe the case. However, their work is not general, from our point of view, in some cases. The most deficient aspect of their work is that they do not consider the decay rate of lower levels coherency and the detuning from resonances which have important effects on the propagation and storage of light in the atomic media. It has caused their theory to be ideal and inexact. In this paper, we try to extend the theory of quantum light memory which is developed previously by M. Fleischhauer et al. to a more general and clear quantum mechanically for slow light and light storage in atomic ensemble with considering all decay rates and detuning.

The organization of this paper is as follows. In section II, quantum mechanical model to describe slow light and light storage is presented. In this section after introducing the mathematical model, two subsection including low intensity limit and small variations and adiabatic passage limit are discussed. The result and discussion is presented in section III. Also, in this section we have two subsections as resonance and off resonance conditions. Numerical simulated result is given in section IV. Finally, the paper is ended with a short conclusion.

## II. Quantum mechanical model to describe slow light and light storage-

Here we further develop the theory presented in [16]. In this case, $\Lambda$ type three level atoms are considered which is demonstrated in Fig. (1). Probe field couples the two $|a\rangle$ and $|b\rangle$ atomic levels together and the respected detune is defined as $\omega_{ab} - \nu_p = \Delta + \Delta_p$. Also the control (coupling) field couples the two $|a\rangle$ and $|c\rangle$ levels with a detuning from resonance ($\omega_{ac} - \nu_c = \Delta$), where $\nu_\mu$ are related to the probe and the control field carrier frequencies and $\omega_{\alpha\beta}$ are the resonance frequencies of corresponding levels. $\Delta$ and $\Delta_p$ are defined as one and two photon detuning respectively.

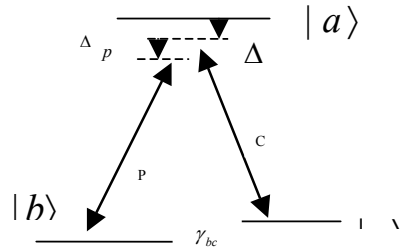

Fig. (1): Schematics of $\Lambda$ type three level atoms

The probe field $\hat{E}(z,t)$ can be defined as follows [16]

$$\hat{E}(z,t) = \sqrt{\frac{\hbar v}{2\epsilon_0 V}} \hat{\mathcal{E}}(z,t) \times e^{i\frac{v_p}{c}(z-ct)} \quad (1)$$

In this relation $\hat{\mathcal{E}}(z,t)$ is the slowly varying annihilation operator (dimensionless field operator) that corresponds to the envelope of probe field. V is the quantization volume of field that can be chosen to be equal to the volume of memory cell. The atomic operator for the atom j is defined as

$$\hat{\sigma}^j_{\alpha\beta} \stackrel{\Delta}{=} |\alpha_j\rangle\langle\beta_j| \quad (2)$$

In this relation $|\alpha_j\rangle$ and $|\beta_j\rangle$ are the Heisenberg Picture base atomic Kets (States) for atom number j. We can divide the memory cell to sections so that atomic operator does not change on them and every division is characterized by coordinate z. By this means one can define the collective (continuum) atomic operators as [16,26]

$$\hat{\sigma}_{\alpha\beta}(z,t) \stackrel{\Delta}{=} \frac{1}{N_z} \sum_{j=1}^{N_z} \hat{\sigma}^j_{\alpha\beta}, \quad (3)$$

where $N_z$ is the number of atoms in the division z. For our purposes, it is easier to work with slowly varying collective atomic operators that are defined as

$$\hat{\tilde{\sigma}}_{\alpha\beta}(z,t) = \hat{\sigma}_{\alpha\beta}(z,t) e^{i\frac{\omega_{\alpha\beta}}{c}(z-ct)}. \quad (4)$$

With considering these operators the interaction Hamiltonian in interaction picture can be written as [27]

$$\hat{H}_I = -N \int \frac{dz}{L} (\hbar g \hat{\mathcal{E}}(z,t) \hat{\tilde{\sigma}}_{ab}(z,t) e^{i(\Delta+\Delta_p)t}$$
$$+ \frac{\hbar}{2} \Omega \hat{\tilde{\sigma}}_{ac}(z,t) e^{i\Delta t}) + h.c., \quad (5)$$

where N is the total number of atoms in the memory cell, L is the length of the cell, and g is the vacuum Rabi frequency that is given by $g = \frac{\vec{\wp}_{ij}\cdot\vec{\varepsilon}\sqrt{\frac{\hbar v_k}{2\epsilon_0 V}}}{\hbar}$ and is related to atom field coupling strength in a given interaction system. Also, $\vec{\wp}_{ij}$ is the electric dipole moment corresponding to the two levels i and j, and $\vec{\varepsilon}$ is the field polarization. One can find equations of motion for the atomic and field operators by substituting the above Hamiltonian in the Heisenberg-Langevin equations [26-29] as

$$[\frac{\partial}{\partial t} + c\frac{\partial}{\partial z}] \hat{\mathcal{E}}(z,t) = igN\hat{\tilde{\sigma}}_{ba}(z,t) \quad (6)$$

$$\frac{\partial}{\partial t} \hat{\tilde{\sigma}}_{bc}(z,t) = -(i\Delta_p + \gamma_{bc})\hat{\tilde{\sigma}}_{bc} - ig\hat{\mathcal{E}}(z,t)\hat{\tilde{\sigma}}_{ac}$$
$$+ i\Omega^* \hat{\tilde{\sigma}}_{ba} + \hat{F}_{bc}(z,t) \quad (7a)$$

$$\frac{\partial}{\partial t} \hat{\tilde{\sigma}}_{ba}(z,t) = -(i(\Delta+\Delta_p) + \gamma_{ba})\hat{\tilde{\sigma}}_{ba}$$
$$+ ig\hat{\mathcal{E}}(z,t)(\hat{\tilde{\sigma}}_{bb} - \hat{\tilde{\sigma}}_{aa}) + i\Omega\hat{\tilde{\sigma}}_{bc} + \hat{F}_{ba}(z,t) \quad (7b)$$

$$\frac{\partial}{\partial t} \hat{\tilde{\sigma}}_{ca}(z,t) = -(i\Delta + \gamma_{ca})\hat{\tilde{\sigma}}_{ca} + i\Omega(\hat{\tilde{\sigma}}_{cc} - \hat{\tilde{\sigma}}_{aa})$$
$$+ ig\hat{\mathcal{E}}(z,t)\hat{\tilde{\sigma}}_{cb} + \hat{F}_{ca}(z,t) \quad (7c)$$

$$\frac{\partial}{\partial t} \hat{\tilde{\sigma}}_{aa}(z,t) = -\gamma_a \hat{\tilde{\sigma}}_{aa} - ig[\hat{\mathcal{E}}^+(z,t)\hat{\tilde{\sigma}}_{ba} - H.a.]$$
$$- i[\Omega^* \hat{\tilde{\sigma}}_{ca} - H.a.] + \hat{F}_a(z,t) \quad (7d)$$

$$\frac{\partial}{\partial t} \hat{\tilde{\sigma}}_{bb}(z,t) = \gamma \hat{\tilde{\sigma}}_{aa} + \gamma'' \hat{\tilde{\sigma}}_{cc}$$
$$+ ig[\hat{\mathcal{E}}^+(z,t)\hat{\tilde{\sigma}}_{ba} - H.a.] + \hat{F}_b(z,t) \quad (7e)$$

$$\frac{\partial}{\partial t} \hat{\tilde{\sigma}}_{cc}(z,t) = \gamma' \hat{\tilde{\sigma}}_{aa} - \gamma'' \hat{\tilde{\sigma}}_{cc} + i[\Omega^* \hat{\tilde{\sigma}}_{ca} - H.a.]$$
$$+ \hat{F}_c(z,t) \quad (7f)$$

The sign ( + ) on operators is the Dagger sign that correspond to Hermitian conjugate of the operators. $\gamma_{\alpha\alpha}$ and $\gamma_{\alpha\beta}$ are the population decay rate of level $\alpha$ and the coherency decay rate of levels $\alpha$ and $\beta$ respectively. $\Omega$ is defined the Rabi frequency of control field that is given by $\Omega = \vec{\wp}_{ac}\cdot\vec{E}_c/\hbar$, where $\vec{E}_c$ is amplitude of the control field. $\hat{F}_\alpha$, $\hat{F}_{\alpha\beta}$ are $\delta$ correlated Langevin noise operators that are caused by reservoir noisy fluctuations (Vacuum Modes) [16,26,28]. In the above equations we see the $\gamma_{bc}$, $\Delta$, $\Delta_p$ terms in which in the main reference [16] they ignored. These parameters, as we will show, especially $\gamma_{bc}$ have considerable effects on the memory behavior. The present equations are a set of coupled differential equations and solving them are difficult. Therefore, we use some approximations to minimize these equations.

**A. Low intensity limit-** For the first approximation, we assume low intensity approximation in which the probe field is very weak compared to the control field. With this approximation, one can consider $\hat{\mathcal{E}}$ as perturbation in above equations and can reach to below relations in the first order of approximation [16,26].

$$<\hat{\tilde{\sigma}}_{bb}(z,t)> \cong 1$$
$$<\hat{\tilde{\sigma}}_{aa}(z,t)>, <\hat{\tilde{\sigma}}_{cc}(z,t)>, <\hat{\tilde{\sigma}}_{ac}(z,t)> \cong 0 \quad (8)$$
$$<\hat{\tilde{\sigma}}_{ba}(z,t)>, <\hat{\tilde{\sigma}}_{bc}(z,t)> \neq 0....(small)$$

The Eq. (7) can then be reduced to the following equations as

$$\frac{\partial}{\partial t} \hat{\tilde{\sigma}}_{bc} = -(i\Delta_p + \gamma_{bc})\hat{\tilde{\sigma}}_{bc} + i\Omega^* \hat{\tilde{\sigma}}_{ba} + \hat{F}_{bc} \quad (9a)$$

$$\frac{\partial}{\partial t} \hat{\tilde{\sigma}}_{ba} = -(i(\Delta+\Delta_p) + \gamma_{ba})\hat{\tilde{\sigma}}_{ba} + ig\hat{\mathcal{E}}$$
$$+ i\Omega\hat{\tilde{\sigma}}_{bc} + \hat{F}_{ba} \quad (9b)$$

The explanation to neglecting some operators in Eq. (7) with respect for Eq. (8) that resulted to Eq. (9) is that most of population in the low intensity approximation is in the state $|b\rangle$ and state of the atomic system tends to $|b\rangle$. Therefore operator $\hat{\tilde{\sigma}}_{bb}$ acts as an identity operator when acting on $|b\rangle$ and operators, $\hat{\tilde{\sigma}}_{aa}, \hat{\tilde{\sigma}}_{cc}$ and $\hat{\tilde{\sigma}}_{ca}$ acting on $|b\rangle$ will result in zero. In addition, the resulting terms $\hat{\mathcal{E}}_{(z,t)}\hat{\tilde{\sigma}}_{cb}, \hat{\mathcal{E}}^+_{(z,t)}\hat{\tilde{\sigma}}_{ba}$, and etc. can be neglected because $\langle\hat{\varepsilon}\rangle, \langle\hat{\tilde{\sigma}}_{ba}\rangle$ and $\langle\hat{\tilde{\sigma}}_{bc}\rangle$ are small. Eq. (9) can be rewritten in the following form as

$$\hat{\tilde{\sigma}}_{ba} = -\frac{i}{\Omega^*}[(i\Delta_p + \gamma_{bc} + \frac{\partial}{\partial t})\hat{\tilde{\sigma}}_{bc} - \hat{F}_{bc}] \quad (10a)$$

$$\hat{\tilde{\sigma}}_{bc} = -\frac{i}{\Omega}(i(\Delta + \Delta_p) + \gamma_{ba} + \frac{\partial}{\partial t})\hat{\tilde{\sigma}}_{ba}$$
$$-\frac{g}{\Omega}\hat{\mathcal{E}} + \frac{i}{\Omega}\hat{F}_{ba}. \quad (10b)$$

For simplicity and clarity of equations and solutions, field-atomic operators are converted to Dark and Bright state equations. The Dark and Bright states are defined respectively as [16]

$$\hat{\Psi}(z,t) \stackrel{\Delta}{=} \cos\theta \times \hat{\mathcal{E}}(z,t) - \sqrt{N}\sin\theta \times \hat{\tilde{\sigma}}_{bc}(z,t),$$
$$\hat{\Phi}(z,t) \stackrel{\Delta}{=} \sin\theta \times \hat{\mathcal{E}}(z,t) + \sqrt{N}\cos\theta \times \hat{\tilde{\sigma}}_{bc}(z,t) \quad (11)$$

where $\hat{\Psi}(z,t)$ is named the Polariton operator that is a superposition of field and atomic state and, as we will see, defines the propagation and storage of Information in the medium. Also, $\theta$ is the control field strength parameter and defined as

$$\tan\theta \stackrel{\Delta}{=} \frac{g\sqrt{N}}{\Omega}. \quad (12)$$

Other relation for $\theta$ are defined as

$$\cos\theta = \frac{\Omega}{\sqrt{\Omega^2 + g^2 N}},$$
$$\sin\theta = \frac{g\sqrt{N}}{\sqrt{\Omega^2 + g^2 N}}. \quad (13)$$

In the above given equations, $\theta$ is a function of time ($\theta(t)$). When the control field is strong enough $\theta$ tends to zero and when the control field is weak or is turned off $\theta$ tends to $\pi/2$. It can easily be verified that expressions for $\hat{\mathcal{E}}(z,t)$ and $\hat{\tilde{\sigma}}_{bc}(z,t)$ are as follows

$$\hat{\tilde{\sigma}}_{bc} = -\frac{1}{\sqrt{N}}(\sin\theta \times \hat{\Psi} - \cos\theta \times \hat{\Phi}), \quad (14a)$$

$$\hat{\mathcal{E}} = \cos\theta \times \hat{\Psi} + \sin\theta \times \hat{\Phi}. \quad (14b)$$

We now get back to the field equation of motion (Eq. (6)) to derive differential equation for $\hat{\Psi}$ and $\hat{\Phi}$. By substituting Eq. (10a) in Eq. (6) we obtain the following equations as

$$[\frac{\partial}{\partial t} + c\frac{\partial}{\partial z}]\hat{\mathcal{E}}(z,t) = igN(-\frac{i}{\Omega^*}[(i\Delta_p + \gamma_{bc} + \frac{\partial}{\partial t})$$
$$\times \hat{\tilde{\sigma}}_{bc} - \hat{F}_{bc}]). \quad (15)$$

By substituting Eq. (14) into the Eq. (15) and doing some mathematical manipulation, we obtain a differential equation in terms of $\hat{\Psi}$ and $\hat{\Phi}$ as

$$\frac{\partial}{\partial t}\hat{\Psi} + c\cos^2\theta\frac{\partial}{\partial z}\hat{\Psi} + (i\Delta_p + \gamma_{bc})\sin^2\theta \times \hat{\Psi}$$
$$= -\hat{\Phi}\dot{\theta} - c\sin\theta\cos\theta\frac{\partial}{\partial z}\hat{\Phi} + (i\Delta_p + \gamma_{bc})$$
$$\times \sin\theta\cos\theta \times \hat{\Phi} - \frac{gN}{\Omega}\hat{F}_{bc}. \quad (16)$$

In deriving the above equation the following assumption are performed.

$$\frac{\partial}{\partial z}\Omega = 0 \Rightarrow \frac{\partial}{\partial z}\theta = 0 \quad (17)$$

This assumption is reasonable because in low intensity approximation, most of the population is in the ground state $|b\rangle$ and therefore, velocity of control pulse is about the speed of light in vacuum. In addition, we have assumed $\Omega$ to be real. Substituting Eq. (14) in Eq. (10b) and using Eq. (10a) yields to another equation For $\hat{\Psi}$ and $\hat{\Phi}$. After doing some manipulation we obtain the following equation as

$$\hat{\Phi} = \frac{\sin\theta}{g^2 N}(i(\Delta + \Delta_p) + \gamma_{ba} + \frac{\partial}{\partial t})(\tan\theta(i\Delta_p + \gamma_{bc} + \frac{\partial}{\partial t}))$$
$$\times (\sin\theta \times \hat{\Psi} - \cos\theta \times \hat{\Phi}) + i\frac{\sin\theta}{g}\hat{F}_{ba}. \quad (18)$$

Eqs. (16,18) are the two general equations to describe the propagation of $\hat{\Psi}$ and $\hat{\Phi}$ in low intensity limit.

**B. Small variations and Adiabatic passage limit-** In order to achieve more simple equations, we assume adiabatic passage limit, which means time variations are small, so that the system have enough time to set itself in Dark state. The conditions for adiabatic passage limit is discussed so far [5,16,31,32]

$$L_p \gg \sqrt{\frac{\gamma_{ba}cL}{g^2 N}} \quad (19a)$$

$$T_r \gg \frac{\gamma_{ba}}{g^2 N}\frac{v_{g0}}{c}, \quad (19b)$$

where $L_p$ is the length of pulse in the medium and L is the total length of memory cell. $T_r$ is characteristic time corresponding to turn on and turn off times of the control field. $v_{g0}$ is the initial group velocity of probe pulse after entering the medium. Eq. (19a) corresponds to adiabatic propagation of light pulse in the medium and it means that the bandwidth of input pulse must be small compared to the transparency window of medium. Eq(19b) corresponds to adiabatic rotation of $\theta$ (turning on and off the control field) that is usually fulfilled in practical situations. In adiabatic limit, Langevin noise operators are negligible because they are $\delta$ correlated [16]. In order to apply the adiabatic passage condition to propagation equations, an adiabatic parameter is introduced as

$$\epsilon \overset{\Delta}{=} \frac{1}{g\sqrt{NT}} \quad (20)$$

where T is a characteristic time corresponding to the probe pulse duration and turn-off and turn on times. We can imagine $\frac{1}{T} \approx \frac{\partial}{\partial t}$ and then replace $\frac{\partial}{\partial t}$ by ($g\sqrt{N}\epsilon$) in Eq. (18). In the zero order of $\epsilon$ ($\epsilon = 0$) that corresponds to adiabatic passage limit, we reach to a simple relation between $\hat{\Psi}$ and $\hat{\Phi}$ as below

$$\hat{\Phi} = \frac{(i(\Delta+\Delta_p)+\gamma_{ba})(i\Delta_p+\gamma_{bc})\tan\theta\sin^2\theta}{g^2 N + (i(\Delta+\Delta_p)+\gamma_{ba})(i\Delta_p+\gamma_{bc})\sin^2\theta}\hat{\Psi}. \quad (21)$$

From this equation it can be inferred that when $\Delta_p$ or $\gamma_{bc}$ is not equal to zero, there will be a population in the bright state ($\hat{\Phi} \neq 0$) that causes a decay of input pulse (information). Consider that in previous work [16], they have obtained $\hat{\Phi} = 0$ in this limit that is because of ignoring $\Delta_p$ or $\gamma_{bc}$. As we will infer these parameters, have considerable effects on the propagation and storage. If we substitute Eq. (21) in Eq. (14) we get relations for atomic and field operators as a function of $\hat{\Psi}$ as below

$$\hat{\tilde{\sigma}}_{bc} = -\frac{1}{\sqrt{N}}(\sin\theta - \cos\theta$$

$$\times \frac{(i(\Delta+\Delta_p)+\gamma_{ba})(i\Delta_p+\gamma_{bc})\tan\theta\sin^2\theta}{g^2 N + (i(\Delta+\Delta_p)+\gamma_{ba})(i\Delta_p+\gamma_{bc})\sin^2\theta})\hat{\Psi} \quad (22a)$$

$$\hat{\mathcal{E}} = (\cos\theta + \sin\theta$$

$$\times \frac{(i(\Delta+\Delta_p)+\gamma_{ba})(i\Delta_p+\gamma_{bc})\tan\theta\sin^2\theta}{g^2 N + (i(\Delta+\Delta_p)+\gamma_{ba})(i\Delta_p+\gamma_{bc})\sin^2\theta})\hat{\Psi}. \quad (22b)$$

We see that if the equation for $\hat{\Psi}$ is known we can get easily the probe field in any time. It can be inferred from Eqs. (21,22b) that when we turn the control field, the light field does not tend to zero, instead its amplitude may be increase because of presence of $\tan\theta$ in Eq. (22b) (As it is also seen in numerical calculations). This result is in contradiction with results of Ref. [16] where they have obtained zero value for the light field when control field is turned off. Because the coefficients are only functions of time, if we obtain the propagation of $\hat{\Psi}$, the same behavior applies for the field and we can obtain it. Therefore, we define $\hat{\Psi}$ as the *information pulse* that can be totally the field or totally the atomic excitation, according to the strength of control field. Information pulse contains the whole of information that is stored. We substitute Eq. (21) into Eq. (16) to reach a differential equation for only $\hat{\Psi}$ as below

---

$$\frac{\partial}{\partial t}\hat{\Psi} + c(\cos^2\theta + B_0)\frac{\partial}{\partial z}\hat{\Psi} + [(i\Delta_p + \gamma_{bc})\sin^2\theta + A_0]\hat{\Psi} = 0 \quad (23)$$

After some manipulation, we obtain relations for $A_0$ and $B_0$ as below

$$A_0 = \frac{(i(\Delta+\Delta_p)+\gamma_{ba})(i\Delta_p+\gamma_{bc})\tan\theta\sin^2\theta \times \dot{\theta} - (i(\Delta+\Delta_p)+\gamma_{ba})(i\Delta_p+\gamma_{bc})^2\sin^4\theta}{g^2 N + (i(\Delta+\Delta_p)+\gamma_{ba})(i\Delta_p+\gamma_{bc})\sin^2\theta}, \quad (24a)$$

$$B_0 = \frac{(i(\Delta+\Delta_p)+\gamma_{ba})(i\Delta_p+\gamma_{bc})\sin^4\theta}{g^2 N + (i(\Delta+\Delta_p)+\gamma_{ba})(i\Delta_p+\gamma_{bc})\sin^2}. \quad (24b)$$

---

As we will discuss later, $A_0$ causes a loss and a phase shift of input pulse. $B_0$ will create a modification of group velocity of Information and light pulses. Therefore, when control field is turned off ($\Omega = 0$), group velocity of Information pulse does not become zero in general. We have calculated this minimum velocity in present paper. In addition $B_0$ causes a k-dependent loss (amplification) which results in dispersion and distortion of the light pulse. To obtain the propagation of Information pulse and therefore that of light pulse in the medium Eq. (23) that is a partial differential equation should be solved. Our presented method is quantum mechanically and $\hat{\Psi}$ is an operator in the above equation. Therefore Eq. (23) governs the propagation of any quantum state of input probe pulse and can be used to study storage of any quantum state in the memory. Therefore, the title "quantum memory" for this type of storage device is justified. To solve the present differential equation and to analyze further the propagation, one can use Fourier Transformation (F.T.) method. It should be considered that $A_0, B_0$ and other coefficients in Eq. (23) are only functions of time (t), so that we can easily get Fourier transform of Eq. (23) with respect to z (space). In addition, one can consider that all coefficients are scalars and Fourier Transformation is just an integral transformation, Therefore, we can extend the F.T. theory to the operators ($\hat{\Psi}$) and get the F.T. of Eq. (23) with considering $\hat{\Psi}$ as an operator. We consider the Fourier transform of $\hat{\Psi}$ as $\hat{\tilde{\Psi}}$ that is also an operator. We can also verify that the differentiation and shifting property of F.T. and all other operations that we will use later is valid for operators. Therefore, our treatment will be ever quantum mechanically. Only when we use the numerical solution to propagation, our treatment becomes classical. Using Fourier Transformation we have

$$\hat{\tilde{\Psi}}(k,t) = \frac{1}{2\pi}\int \hat{\Psi}(z,t)e^{-ikz}dz. \quad (25)$$

In which $\hat{\tilde{\Psi}}$ is the Fourier Transform of $\hat{\Psi}$ in momentum domain. By applying F.T. to Eq. (23), we obtain an ordinary differential equation as

$$\frac{\partial}{\partial t}\hat{\tilde{\Psi}}(k,t) + [(i\Delta_p + \gamma_{bc})\sin^2\theta + A_0$$
$$+ ikc(\cos^2\theta + B_0)]\hat{\tilde{\Psi}}(k,t) = 0. \quad (26)$$

This equation can be solved by simple integration and the solution is

$$\hat{\tilde{\Psi}}(k,t) = \hat{\tilde{\Psi}}(k,0)\exp[-\int_0^t [(i\Delta_p + \gamma_{bc})\sin^2\theta$$
$$+ A_0 + ikc(\cos^2\theta + B_0)]dt]. \quad (27)$$

In which $\hat{\tilde{\Psi}}(k,0)$ is the F.T. of input information pulse (Polariton) at t=0 (Corresponding to input light pulse with Eqs. (11-14)). Integrating the above equation is difficult because $\theta$ is usually a complicated function of time when we are interested in various profiles of turning on and off the control field that is determined using $\theta(t)$ profile. Therefore, we should use the numerical methods to obtain output field from a given input field. In numerical calculation we switch to scalar (classical) values $\mathcal{E}, \Psi$ and $\Phi$ that are expectation values of operators $\hat{\mathcal{E}}, \hat{\Psi}, \hat{\Phi}$ in the system. We calculate the scalar form of integral in Eq. (27), then get its Inverse Fourier Transform and then insert it in scalar counterpart of Eq. (22) to get $\mathcal{E}(z,t)$ in any time and location.

### III. Results and Discussion

Before presenting the numerical results, we return to Eq. (27) to do some more analytical analysis on it. Eq. (27) can be rewritten in the following form

$$\hat{\tilde{\Psi}}(k,t) = \hat{\tilde{\Psi}}(k,0)\exp[-\int_0^t [\alpha_1 + i\beta + k\alpha_2 + ik\mathbf{V}_g]dt] \quad (28)$$

We can calculate $\alpha_1, \alpha_2, \beta, v_g$, from inserting expressions for $A_0, B_0$ into the Eq. (27). All of these coefficients are only functions of time. Eq. (28) is a valuable equation to understand the behavior of quantum memory and to predict the output light. We can now interpret every coefficient by considering Eq. (28). The $\alpha_1$ term determines the decay rate of the information pulse in every time and is the same for all k's, so that it will not cause any dispersion or distortion of the information pulse, so that it will not cause to loss of information. It only causes a decay of total pulse by the rate of $\alpha_1(t)$. The $\beta$ term corresponds to a phase shift of total information pulse. The $\alpha_2$ term is a k-dependent loss (Amplification) of information pulse that will cause to dispersion and distortion of information and light pulse, so that has the worst effect on storing information. $\mathbf{V}_g$ is the velocity of information pulse in every time as we can infer it from the shifting property of F.T.

**Small detuning-High atomic density limit-** If we restrict ourselves to small detuning and high atomic densities, we can get our equations simpler. Therefore, we assume below condition

$$g^2 N \gg |(i(\Delta + \Delta_p) + \gamma_{ba})(i\Delta_p + \gamma_{bc})| \quad (29)$$

Considering the typical range for parameters in the above equation, one finds that the present condition is usually fulfilled. In this case the denominator in $A_0$ and $B_0$ reduces to $g^2 N$. We write the reduced $A_0$ and $B_0$ in their imaginary and real parts as below

----

$$A_0 = \frac{i\sin^2\theta}{g^2 N}[((\Delta+\Delta_p)\gamma_{bc} + \Delta_p \gamma_{ba})(\tan\theta \times \dot\theta - \gamma_{bc}\sin^2\theta) - (\gamma_{bc}\gamma_{ba} - \Delta_p(\Delta+\Delta_p))(\Delta_p \sin^2\theta)]$$
$$+ \frac{\sin^2\theta}{g^2 N}[(\gamma_{bc}\gamma_{ba} - \Delta_p(\Delta+\Delta_p))(\tan\theta \times \dot\theta - \gamma_{bc}\sin^2\theta) + ((\Delta+\Delta_p)\gamma_{bc} + \Delta_p\gamma_{ba})(\Delta_p \sin^2\theta)], \quad (30a)$$

$$B_0 = [i(\Delta+\Delta_p)\gamma_{bc} + i\Delta_p\gamma_{ba} + \gamma_{bc}\gamma_{ba} - \Delta_p(\Delta+\Delta_p)]\frac{\sin^4\theta}{g^2 N}. \quad (30b)$$

We then calculate the expressions for $\alpha_1, \alpha_2, \beta, v_g$ by considering Eqs. (27, 28, 30) as below

$$\alpha_1 = \gamma_{bc}\sin^2\theta + \frac{\sin^2\theta}{g^2 N}[(\gamma_{bc}\gamma_{ba} - \Delta_p(\Delta+\Delta_p))(\tan\theta \times \dot\theta - \gamma_{bc}\sin^2\theta) + ((\Delta+\Delta_p)\gamma_{bc} + \Delta_p\gamma_{ba})(\Delta_p\sin^2\theta)]. \quad (31a)$$

$$\alpha_2 = -c((\Delta+\Delta_p)\gamma_{bc} + \Delta_p\gamma_{ba})\frac{\sin^4\theta}{g^2 N}, \quad (31b)$$

$$\beta = \Delta_p \sin^2\theta + \frac{\sin^2\theta}{g^2 N}[((\Delta+\Delta_p)\gamma_{bc} + \Delta_p\gamma_{ba})(\tan\theta \times \dot\theta - \gamma_{bc}\sin^2\theta) - (\gamma_{bc}\gamma_{ba} - \Delta_p(\Delta+\Delta_p))(\Delta_p\sin^2\theta)], \quad (31c)$$

$$v_g = c(\cos^2\theta + (\gamma_{bc}\gamma_{ba} - \Delta_p(\Delta+\Delta_p))\frac{\sin^4\theta}{g^2 N}). \quad (31d)$$

----

These are valuable relations that we can determine the properties for propagation and storage of information in the medium.

**a) Resonance condition-** We now derive the above equations for the case of zero detuning ($\Delta = \Delta_p = 0$) and result are given as

$$\alpha_1 = \gamma_{bc}\sin^2\theta + \gamma_{bc}\gamma_{ba}(\tan\theta \times \dot\theta - \gamma_{bc}\sin^2\theta)\frac{\sin^2\theta}{g^2 N}, \quad (32a)$$

$$\alpha_2 = 0, \quad (32b)$$

$$\beta = 0, \quad (32c)$$

$$v_g = c(\cos^2\theta + \gamma_{bc}\gamma_{ba}\frac{\sin^4\theta}{g^2 N}). \quad (32d)$$

Usually the second term in $v_g$ is small compared to the first term, unless the control field is off where the first term tends to zero. From the above equation, we can derive that when we turn off the control field ($\theta$ tends to $\pi/2$ and $\cos\theta \to 0$), we have the minimum velocity for information pulse (light pulse) in resonance case as

$$v_{g\min} = c\frac{\gamma_{bc}\gamma_{ba}}{g^2 N} \neq 0. \quad (33)$$

Consider that we get a non-vanishing velocity for information pulse and it is because there is a non-vanishing $\gamma_{bc}$ in the system. This equation can be regarded as a limit to the maximum value of storage time for a given length of the medium. We see a very good agreement between the Eq. (33) and the numerical results that will be presented later. In experimental high atomic density conditions, we usually have $g^2 N \gg \gamma_{ba}\gamma_{bc}$ so we can neglect the third term in $\alpha_1$, so we get to an even simpler expression for $\alpha_1$ as

$$\alpha_1 = (\gamma_{bc} + \frac{\gamma_{bc}\gamma_{ba}}{g^2 N}\tan\theta \times \dot\theta)\sin^2\theta. \quad (34)$$

**Slow light conditions-** If we set the initial speed of light pulse in the medium (when the control field is on) to be very smaller than the speed of light in vacuum, then $\theta$ will be very close to $\pi/2$ and we can assume $\sin\theta$ equal to unity and get to

$$\alpha_1 = \gamma_{bc} + \frac{\gamma_{bc}\gamma_{ba}}{g^2 N}\tan\theta \times \dot\theta. \quad (35)$$

We see that for $\dot\theta = 0$ i.e. when the control field strength is constant, damping rate reduces to very simple relation

$$\alpha_1 = \gamma_{bc}. \quad (36)$$

That is equal to the damping rate of lower levels coherency. One can verify that Eq. (36) is identical to the numerical results (Fig. (3,4)). For $\dot\theta \neq 0$, $\dot\theta$ term portion to damping is small when control field has still a considerable value. i.e. when $\frac{\gamma_{ba}}{g^2 N}\tan\theta \times \dot\theta \ll 1$. ($\tan\theta$ is not very large).

The $\dot\theta$ term will cause an additional decay rate only when $\Omega$ is changing and is very small. Therefore, it effect is considerable in very small times during the turn on and turn off of the control field and we can neglect its effect with some approximation on the overall damping of information pulse. Therefore, we can use an approximate, but very simple relation for the output probe pulse when it is coming out of the memory cell after a storage time of $T_0$ as below

$$\hat{\mathcal{E}}(t)|_{z_{out}} = \hat{\mathcal{E}}_0(t - T_0)|_{z_{in}} e^{-\gamma_{bc} T_0}. \quad (37)$$

Also $z_{out}$ and $T_0$ are related to each other by the equation below

$$z_{out} - z_{in} = \int_0^{T_0} v_g \, dt. \quad (38)$$

We see a good agreement between the results of numerical calculations (Fig. (3)) and the results obtained by Eq. (37) for damping of information pulse. Therefore, we can claim that approximation used for Eq. (37) is a proper approximation for usual cases. Of course, if we want to calculate the damping exactly, we can use the below equation to obtain the output field

$$\hat{\mathcal{E}}(t)|_{z_{out}} = \hat{\mathcal{E}}_0(t - T_0)|_{z_{in}} e^{-\int_0^{T_0} \alpha_1 dt}. \quad (39)$$

In which $\alpha_1$ is given by Eq. (35). In addition, if we want to calculate the output in a condition other than slow light condition i.e. for the case that our information pulse is propagating with a speed comparable to $c$ in the medium when the control field is on, we can use the Eq. (34) in the Eq. (39) to calculate the overall damping. With considering the above results, we see that if we ignore $\gamma_{bc}$ (that has the typical range of $10^2 - 10^5 \, rad/\sec$) in our relations (as it is done in previous work [16]), we reach to idealistic and incorrect results about propagation and storage of light in the memory medium.

**b) Off- resonance condition-** In the off-resonance case, $\alpha_2, \beta$ becomes nonzero in general. $\beta$ correspond to a phase shift of total in formation pulse that do not considerably affect on destroying stored information. However, nonzero $\alpha_2$, if it is considerable, will cause a k-dependent amplification leading to distortion and dispersion (fast oscillations) of the light pulse (as it is also seen in numerical calculations). In quantum memory, we should minimize any distortion and dispersion because it destroys the stored information. We can see from Eq. (31b), that if $\Delta, \Delta_p$ have different signs, they will reduce effects of each other and even they can be adjusted to $\alpha_2 = 0$. However, it will cause an additional decay rate and increased minimum group velocity that is not desired. In addition, we can set $v_{g\min}$ in Eq. (31d) to be zero, however it will cause a considerable $\alpha_2$ (regarding Eq. (31b)) that destroys completely the stored information. We also try to decrease $\alpha_1$ (the decay rate) compared to resonance case, by setting $\Delta, \Delta_p$ to proper values in Eq. (31a).

However, we see that every attempt to reduce $\alpha_1$ causes in considerable $\alpha_2$ and destroyed information. Therefore, we deduce that any detuning (in small detuning--high atomic density limit, Eq. (29)) is undesired in quantum memory and will result in loss of stored information. So, we should try to set the system to resonance. In addition, we deduce that the system is extremely more sensitive to $\Delta_p$ than $\Delta$ (consider Eq. (31b)) and a small $\Delta_p$ (comparable to $0.01 \frac{g^2 N L_p}{c \gamma_{ba} T_0}$) will cause a complete loss of information. $T_0$ is the storage time and $L_p$ is the length of pulse in the medium The limitations to $\Delta, \Delta_p$ in order to have acceptable (undistorted) output are derived from Eq. (28,31b) as follows

$$\Delta_p \ll 0.01 \frac{g^2 N L_p}{c \gamma_{ba} T_0}, \tag{40a}$$

$$\Delta \ll 0.01 \frac{g^2 N L_p}{c \gamma_{bc} T_0}. \tag{40b}$$

Consider that in our model, we have assumed a single carrier frequency for the probe and coupling fields that are idealistic and impractical (This assumption is common (as it is assumed in previous works), because it causes an enormous simplification in analysis). However, in practical situation, our lasers have a considerable bandwidth in their nature and we cannot reduce their bandwidths more than a specified range. We can consider the present system as a set of probe and coupling fields with different carrier frequencies applied to atomic system that causes to presence and action of various $\Delta, \Delta_p$ on the system. Therefore, there will be an undesired distortion and loss of stored information. We finally deduce that it is necessary to use lasers with very narrow bandwidth to get an acceptable efficiency for quantum storage of light. We can infer from Eq. (40) that our lasers must have narrow Bandwidths also, center frequencies and bandwidths very close to each other as follows

$$|BW_c - BW_p|, |\nu_{0c} - \nu_{0p}| \ll 0.01 \frac{g^2 N L_p}{c \gamma_{ba} T_0}, \tag{41a}$$

$$BW_{c,p} \ll 0.01 \frac{g^2 N L_p}{c \gamma_{bc} T_0}. \tag{41b}$$

In the recent equation, $BW_\mu, \nu_{0\mu}$ are the Bandwidth and the center frequencies of the corresponding lasers.

**IV. Numerical Calculations-** We have calculated the scalar counterpart of integral in Eq. (27) and other procedure numerically to obtain the Information Pulse and light pulse in any time and space. We use the typically reported values for the parameters in this numerical calculation [7-9,23-25]. We set the parameters as follow. $L = 5^{mm}$ (Length of the memory cell); $D = 200^{\mu m}$ (Diameter of the Cell); $\wp_{ca} \approx \wp_{ba} \approx 10^{-29} - 10^{-28}$ C.m (Electric dipole of corresponding levels); $\nu_p = 5 \times 10^{14} Hz \times 2\pi$ (Probe field frequency); $g = 10^6 \, rad/\sec$ (Vacuum Rabi frequency); $N = 10^8 \, atoms$ (Number of atoms in the cell corresponding to the atomic density of $10^{12} cm^{-3}$). We use the input Information pulse as

$$\Psi(z=0,t) = 0.2 \times e^{-(\frac{z}{10^{-3}})^2} \tag{42}$$

That corresponds to a especial Rabi frequency of the probe field ($\Omega_p = 5 \times 10^5 \, rad/\sec$). In addition, we use a profile to turn off and turn on the control field as follows:

$$\theta = Arc \cot[5 \times 10^{-4} \times \{1 - 0.5 \times \tanh(10^5 \times (t - t_1))$$
$$+ 0.5 \times \tanh(10^5 \times (t - t_2))\}] \tag{43}$$

The parameters $t_1, t_2$ are the turn off and turn on times that we set to be $t_1 = 30 \times 10^{-6}$ sec, $t_2 = 125 \times 10^{-6}$ sec. We have shown the $\theta, \Omega$ in the Fig. (2) as a function of time. We have set by above Equation the Rabi frequency of control field (Control field strength) in the "on time" to be about $\Omega_c = 5 \times 10^6$ rad/sec that results in the initial velocity of light in the medium to be about $(75+ v_{g\min})$ m/sec (Slow initial velocity). As one may consider $\theta$ is very close to unity even when the control field is on. In this case, dominant part of Information pulse (Polariton) is the atomic excitations even when the control field is on. Setting the initial velocity of light to be small is reasonable in the practical case to trap the light. If the initial velocity of light is near speed of light it will escape the medium before the control field is turned off and its control would be difficult (as this is done in experimental works [23,24]). We have plotted the velocity of Information pulse in Fig.2 using Eq. (32d).

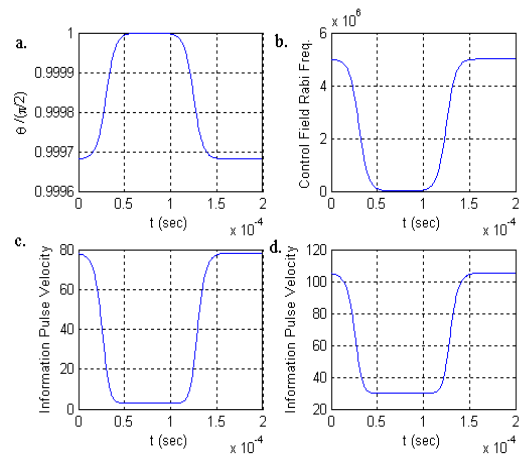

Fig. (2): a) $\theta/_{\pi/2}$ as a function of time. **b)** Rabi frequency of control field as a function of time. c) Group velocity of information pulse for $\gamma_{ba} = 10^8$ $rad/\sec$ . d) Group velocity of information pulse for $\gamma_{ba} = 10^9$ $rad/\sec$ . ( $\gamma_{bc} = 10^4$ $rad/\sec$ , $\Delta, \Delta_p = 0$ .)

**Resonance case-** Fig. (3) shows the numerical results for information pulse propagation and storage for steps of 15 $\mu$ sec and given values of parameters. We can verify that the results in Eqs. (36,37) are identical to the numerical results. Fig. (4) shows the information pulse in storage region where the control field if off. One can easily verify the minimum group velocity (Eq. (33)) and the damping rate (Eq. (36)) is identical to numerical results. We can see the differences between the results of previous works [16] and our results. In previous works, there is no decay for the information pulse (Polariton) in adiabatic passage limit and the information pulse is slowed down to stop when control field is off. Fig. (5) shows the Bright state pulse for the time steps of 15 $\mu$ sec . As it is shown the bright state grows to high values when the control field is off. It is because of presence of $\tan\theta$ in the Eq. (21). Because of the Bright state in off time of control field is very higher than that in other times, in Fig. (5b,5d) the Bright state for other times are very small that can not be seen. Consider that in previous works [16], the Bright state is always zero valued in the adiabatic passage limit. We illustrate a typical initial Bright state at t=0 ( $\gamma_{ba} = 10^8$ , $\gamma_{bc} = 10^4$ $rad/\sec$ ), in Fig. (5a) to compare with Bright states at other times. (Consider the scaling in figures).

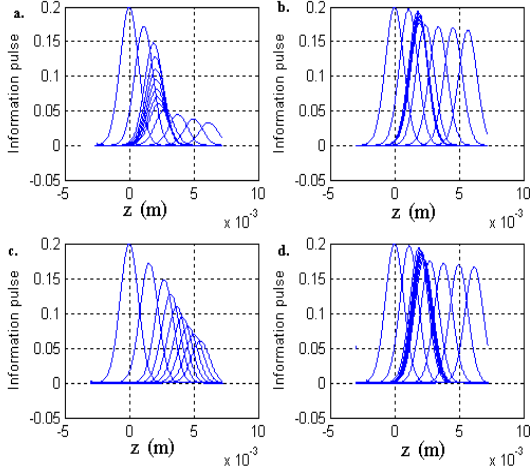

Fig. (3): Information pulse propagation for time steps of 15 $\mu$ sec .a) $\gamma_{ba} = 10^8$ , $\gamma_{bc} = 10^4$ ( $rad/\sec$ ). b) $\gamma_{ba} = 10^8$ , $\gamma_{bc} = 10^3$ ( $rad/\sec$ ). c) $\gamma_{ba} = 10^9$ , $\gamma_{bc} = 10^4$ ( $rad/\sec$ ). d) $\gamma_{ba} = 10^9$ , $\gamma_{bc} = 10^3$ ( $rad/\sec$ ). ( $\Delta, \Delta_p = 0$ ).

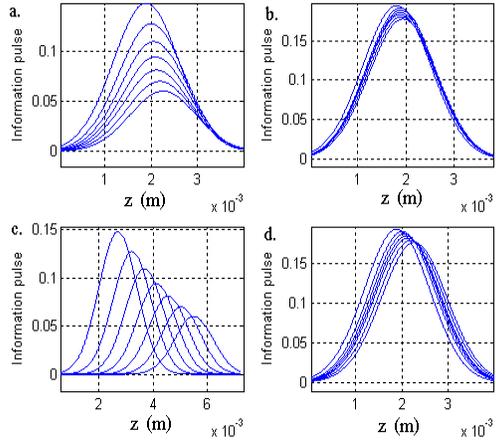

Fig. (4): Information pulse for time steps of $15\mu$ sec in the storage region where the control field is off.
a) $\gamma_{ba} = 10^8$ , $\gamma_{bc} = 10^4$ ( $rad/\sec$ ). b) $\gamma_{ba} = 10^8$ , $\gamma_{bc} = 10^3$ ( $rad/\sec$ ). c) $\gamma_{ba} = 10^9$ , $\gamma_{bc} = 10^4$ ( $rad/\sec$ ). d) $\gamma_{ba} = 10^9$ , $\gamma_{bc} = 10^3$ ( $rad/\sec$ ). ( $\Delta, \Delta_p = 0$ ).

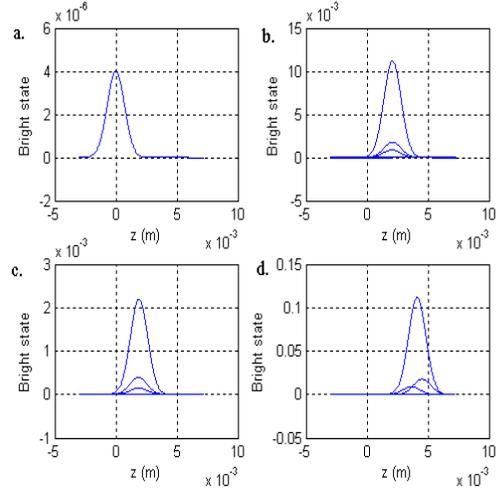

Fig. (5): a) Initial bright state for input pulse with $\gamma_{ba} = 10^8$ , $\gamma_{bc} = 10^4$ $rad/\sec$ . b) Bright state for time steps of $15\mu$ sec for $\gamma_{ba} = 10^8$ , $\gamma_{bc} = 10^4$ $rad/\sec$ . c) Bright state for time steps of $15\mu$ sec and for $\gamma_{ba} = 10^8$ , $\gamma_{bc} = 10^3$ $rad/\sec$ . d) Bright state for time steps of $15\mu$ sec and for $\gamma_{ba} = 10^9$ , $\gamma_{bc} = 10^4$ $rad/\sec$ . ( $\Delta, \Delta_p = 0$ ).

Fig. (6) show the $\mathcal{E}_{(z)}$ (Dimensionless Field envelope) for the time steps of 15 $\mu$ sec in propagation. We see that the velocity for the probe pulse is identical to the information pulse and the information is turned back to the probe field as it was at the input ( $t = 0$ ), when we turn on the control field again. We reach to an interesting result that when the control field is turned off, the probe field grows to very high values compared to its initial value in the medium. We consider that there is a contradiction between the present results and the results of [16] in which the probe field is completely converted to atomic excitations when the control field is turned off. This contradiction is because of ignoring $\gamma_{bc}$ in the main reference.

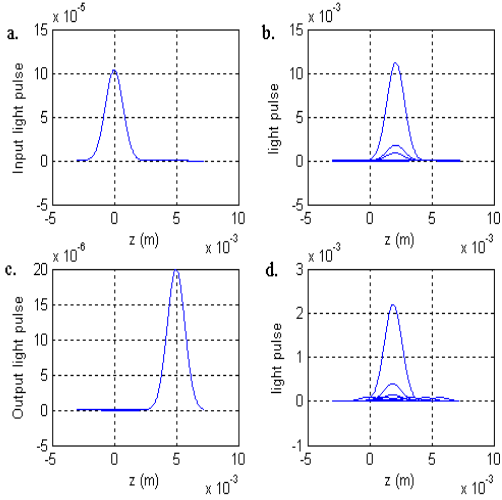

Fig. (6): a) Input light pulse inside medium for $\gamma_{ba} = 10^8$, $\gamma_{bc} = 10^4$ $rad/\sec$. b) light pulse for time steps of $15\mu\sec$ for $\gamma_{ba} = 10^8$, $\gamma_{bc} = 10^4$ $rad/\sec$. c) Output light pulse inside medium for $\gamma_{ba} = 10^8, \gamma_{bc} = 10^4$ $rad/\sec$. d) light pulse for time steps of $15\mu\sec$ and for $\gamma_{ba} = 10^8$, $\gamma_{bc} = 10^3$ $rad/\sec$. ($\Delta, \Delta_p = 0$)

Fig.7 shows the atomic excitation ($\sigma_{bc}$) for various times. We see that it only decays by the rate of $\alpha_1$ and when we turn off the control field, there is no considerable variation (In contrast with results of [16]). This difference is not fundamental and is because we have set the initial velocity of the light pulse to very small value (as it is done in practical reports [23,24]) and dominant part of the information pulse is atomic excitations from the beginning.

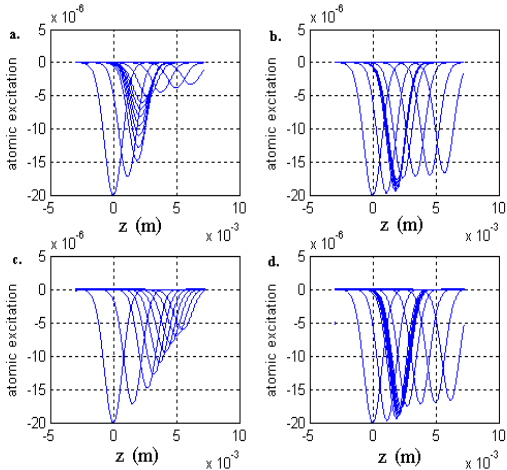

Fig. (7): a) $\sigma_{bc}$ (atomic excitation) propagation for time steps of $15\mu\sec$. a) $\gamma_{ba} = 10^8$, $\gamma_{bc} = 10^4$ ($rad/\sec$).

b) $\gamma_{ba} = 10^8$, $\gamma_{bc} = 10^3$ ($rad/\sec$). c) $\gamma_{ba} = 10^9$, $\gamma_{bc} = 10^4$ ($rad/\sec$). d) $\gamma_{ba} = 10^9$, $\gamma_{bc} = 10^3$ ($rad/\sec$). ($\Delta, \Delta_p = 0$).

**Off-resonance case-** Fig. (8,9) shows the Information pulse at the times $t = 0$ (input) and $t = 165\mu\sec$ (output), for various $\Delta, \Delta_p$'s. We can now examine the condition in Eq. (40) to be correct. We see that when the condition in Eq. (40a) or Eq. (40b) is violated the output information pulse bears fast oscillations and gets destroyed that causes the loss of total information. Upon the limit in Eq. (40), a very small increasing in detuning causes a considerable increase in distortion. As stated before, this is because we have a the k-dependent loss (amplification) $\alpha_2$ in the Eq. (28). With the values given, we see that any attempt to reduce $v_{gmin.}, \alpha_1$ in Eq. (31) with setting $\Delta, \Delta_p$, will cause a complete loss of information. From Fig. (8,9), it can be inferred that if our practical lasers are not well adjusted and have not very narrow bandwidths, we will not be able to have any storage of information. (Consider that Eq. (41) sets a very small and strict limit for the Bandwidth and center frequency of lasers and these values are difficult to achieve by conventional lasers and optical devices). We here refer to [17]. In that paper, they have set a limit to Two-photon detuning $\Delta_p$ (that is referred as $\delta$ there (Eq. (23)). Their limitation is far larger than the limit that we have obtained in Eq. (40) and is not the effective limit because they have not considered the k-dependent loss ($\alpha_2$) in calculating this limit. In addition, their equations are somewhat incorrect because they have used the incorrect signs in their equations (4,6) ([26,27]).

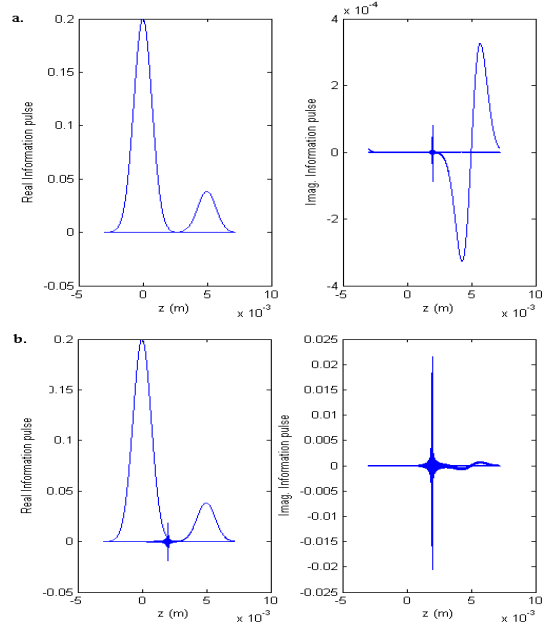

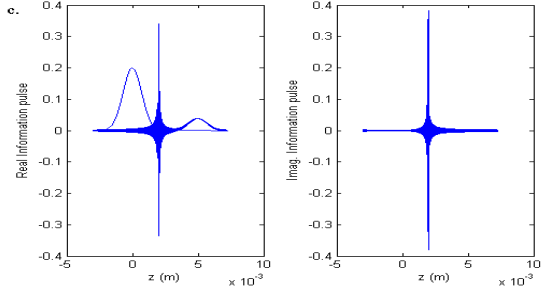

Fig. (8): Real and Imaginary parts of the information pulse at the times $t = 0$ (input) and $t = 165\mu\sec$ (output) plotted with each other for various values of One-Photon detuning.

a) $\Delta = 2\times 10^6 \ rad/\sec$   b) $\Delta = 4\times 10^6 \ rad/\sec$

c) $\Delta = 5\times 10^6 \ rad/\sec$.  ($\Delta_p = 0, \gamma_{ba} = 10^8, \gamma_{bc} = 10^4 \ rad/\sec$).

Then these have caused their following equations (11,12,13, 14) to be inexact (compare with our Eqs. (7,16,18,21,23).

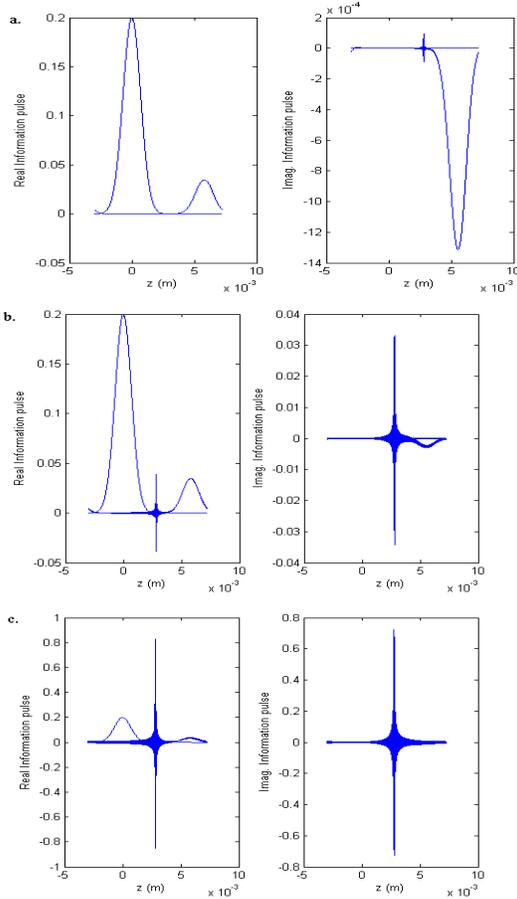

Fig. (9): Real and Imaginary parts of the information pulse at the times $t = 0$ (input) and $t = 165\mu\sec$ (output) plotted with each other for various values of Two-Photon detuning.

a) $\Delta_p = 2\times 10^2 \ rad/\sec$.  b) $\Delta_p = 4\times 10^2 \ rad/\sec$.

c) $\Delta_p = 5\times 10^2 \ rad/\sec$.

($\Delta = 0, \gamma_{ba} = 10^8, \gamma_{bc} = 10^4 \ rad/\sec$).

**V) Conclusion-** In this paper, we further developed the quantum mechanical theory for quantum light memory in the Low intensity limit and small variations and adiabatic passage limit, primarily developed by [16]. We entered the parameters $\gamma_{bc}, \Delta, \Delta_p$ into the formulations. We obtained and explained their effects in a clear form. We analyzed the propagation and storage of the light pulse in the resonance case and obtained the decay rate and the minimum group velocity in this case. In addition, we reached to a non-vanishing light field when the control field is turned off in storage process. We then analyzed the off-resonance case and reached to the result that off-resonance case (In Small detuning and high atomic density limit) has no advantage and can destroy completely the output (stored) information. We obtained a limit to maxi-mum value of detuning in order to maintain the stored information and therefore a limit to band-widths and center frequencies of practical lasers used. We then presented the related numerical results to verify the analytical results and limits.